  \documentclass[aps,pre,superscriptaddress,groupedaddress,12pt]{revtex4}  

\usepackage{graphicx}  
\usepackage{dcolumn}   
\usepackage{bm}        
\usepackage{amssymb}   
\usepackage{amsmath}
\usepackage{color,soul}

\begin{document}

\title{The cost of Bitcoin mining has never really increased} 
\author{Yo-Der Song} 
\author{Tomaso Aste} 
\affiliation{Department of Computer Science \& Centre for Blockchain Technologies, University College London, Gower Street,  WC1E 6EA, London, United Kingdom}


\begin{abstract}
The Bitcoin network is burning a large amount of energy for mining. 
In this paper we estimate the lower bound for the global energy cost for a period of ten years from 2010, taking into account changing oil costs, improvements in hashing technologies and hashing activity. 
Despite a ten-billion-fold increase in hashing activity and a ten-million-fold increase in total energy consumption, we find the cost relative to the volume of transactions has not increased nor decreased since 2010.
This is consistent with the perspective that, in order to keep a the Blockchain system secure from double spending attacks, the proof or work must cost a sizable fraction of the value that can be transferred through the network  \citep{aste2016fair,aste2017blockchain}. 
We estimate that in the Bitcoin network this fraction is of the order of 1\%.\\
{\bf \small Keywords: Bitcoin, mining cost, proof of work, cryptocurrency, blockchain}
\end{abstract}
\maketitle

\section{Introduction}
Bitcoin is a digital currency launched in 2009 by an anonymous inventor or group of inventors under the alias of Satoshi Nakamoto \citep{satoshi}. 
It is the largest cryptocurrency in market capitalization with over 100 billion dollars \citep{bitcoinfirst,bitcoinmarketcap,marketcap}. 
As a decentralized currency, Bitcoin differs from government regulated fiat currencies in that there exists no central authority within the network to verify transactions and prevent frauds and attacks. 
Instead, Bitcoin relies on a highly replicated public ledger, secured by means of a hash chain and validated through community consensus.  
All users can announce a new transaction but such a transaction will be considered valid and included in the ledger only once it is verified by a majority of the network nodes. 
Transactions are written into blocks that are interlocked into a chain by hashes.
Hashing is a one-way function that maps an input of arbitrary length into a string of a fixed number of digits. 
The hash function must guarantee that the output string is (quasi-)uniquely related to the given input (deterministic) and that small changes in the input should cause arbitrarily large changes in the output so that reconstructing the input based on the output is infeasible. 
In the case of Bitcoin, the transactions in the new proposed block and the header of the most recent block is inputted into the SHA-256 hash algorithm, making therefore a chain with unique direction.
Such a chain is at the heart of the Bitcoin security because it makes it difficult to alter the content of a block once subsequent blocks are added to the chain.
In Bitcoin, this cryptographic sealing process through a hash chain is intentionally designed to be computationally intensive by accepting hashes only if the randomly generated hash number is smaller than a given target. 
Therefore the community performs a large number of hashing by modifying a random component of the block content until, by chance, someone finds a `valid' hash that is smaller than the threshold. 
This is called proof of work (PoW) and serves the purpose to determine majority consensus. 
Indeed, in an anonymous distributed system, participants can arbitrarily generate new identities so consensus cannot be accounted in terms of individuals. Rather, it must be accounted in terms of some participation cost demonstrating the commitment of computational power.
In the words of Satoshi Nakamoto, “one CPU one vote” \cite{satoshi}. 

Users are incentivized to participate to the block validation by assigning newly mined Bitcoins to the first who randomly finds a hash with a value smaller than the threshold. 
Presently, this remuneration is 6.25 Bitcoins, corresponding to around USD 60,000 at the current exchange rate (see Table \ref{table:reward}). 
For this reason the hashing process is called `mining'.

Sometimes forks occur in the blockchain when two blocks containing different transactions are attached to the same block. Eventually other blocks are mined and attached to them, forming two branching chains after the fork.
In this case, the longer chain, the one with more cumulative proof of work or hash computations, would be considered as the main chain upon which future blocks are built on. 
Normally a block is considered finally valid after 6 blocks are attached to its chain, which takes approximately one hour.

The Bitcoin proof of work is very costly. 
Technological improvements over the years have made hashing a very efficient operation, consuming at little as 0.03 joules per billion hashes (with specifically-designed Application-Specific Integrated Circuit, ASIC, machines. See Table \ref{table:hardware}). 
This has reduced energy cost per hash by about thirty thousand times during the last ten years.
However, the miners in the Bitcoin network are presently (May 2020) computing nearly $10^{25}$ hashes per day, up over 10 orders of magnitude from the 2010 levels. 
We estimate in this paper that this hashing activity currently corresponds to an energy cost of around 1 million USD per day and around a billion USD over the past year.
In turn, this corresponds a per transaction costs as high as 13 USD in January 2020. 
This cost is not borne by either the sender nor the receiver in a transaction but rather by the miners. 

While a billion a year burned in hashing is definitely a large amount of money that could be seen as a waste of resources, the Bitcoin proof of work is a necessary process for such an anonymous permission-less network to function.
It is indeed required to validate transactions and obtain community consensus to secure the system from attacks.  

One question arises: is this cost fair or could it be lowered?
In  \cite{aste2016fair} made the argument that, at equilibrium, the cost of Bitcoin proof of work should be such to make a double spending attack too expensive to be profitably carried out. 
From this principle, it is relatively straightforward to estimate the fair cost of the proof of work under an ideal equilibrium assumption. 
Let us consider an attacker that owns some amount of Bitcoin and wants to artificially multiply it by spending the same Bitcoin with several different users. This is known as a double spend attack. The attacker will try to double spend the largest amount of Bitcoin possible, but this is limited to the amount normally exchanged within a block (which, we estimate in this paper, is currently around \$10 million). Indeed, a transaction involving a substantially larger sum than the usual will capture unwanted attention from the network. 
Of course, the duplication can be repeated several times both in parallel or serially but, as we shall see shortly, this does not affect the outcomes of the present argument. 
To be successful the attacker must make sure that both the duplicated transactions are validated and this requires the generation of a fork with two blocks containing the double spent transaction attached to the previous block.  If the attacker has sufficient computing power, she can generate two valid hashes to seal the two blocks giving the false impression that both transactions have been verified and validated. However, for a final settlement of the transaction, it is presently considered that one should wait six new blocks to be attached to the chain to make the transaction statistically unlikely to be reverted. The attacker should therefore use her computing power to generate six valid hashes before the double spent transaction might be considered settled. Note that only one of the two forks (the shortest) must be artificially validated by the attacker since the other will be considered valid by the system and can be let to propagate by the other miners. 
Of course, it is quite unrealistic to assume that nobody notices the propagating fork for such a long time, but let’s keep this as a working hypothesis. The artificial propagation of the fork has a cost that is the cost of the proof of work per block times six. 
The attacker will make profits if this cost is inferior to the gain made from duplicated spending. 
In the previous unpublished note by \cite{aste2016fair} the following formula is reported:
\begin{equation}
\label{eqn:fair0}
\text{Equilibrium fair cost of proof of work per block} = \frac{\text{duplicated fraction of the value of a block}}{\text{number of blocks required for settlement}}.
\end{equation}
 We can re-write this formula to formally express the cost of proof of work per day, $C_{t}$, as  
\begin{equation}
\label{eqn:fair}
C_{t} = \frac{p V_{t}}{N}
\end{equation}
where:\\
 $p$ is the duplicated proportion of block transaction volume; \\
 $V_{t}$  is the average  transaction volume on day $t$; \\
 $N$ is the number of blocks required for settlement.

In Eq.\ref{eqn:fair}  $N$ is roughly equal to 6 and the current average volume of transaction is  about $V_t\sim$ 1 billion USD a day but it was only a few thousands dollars a day in 2010.
The value of $p$ must be considerably smaller than one because an attacker will be spotted immediately by the community if she tries to fork with a large double-spent value with operations that involve a significant portion of the entire network activity. 
We must note that this formula is an upper bound for the cost of the proof of work. 
It greatly underestimates the costs of an attack and largely overestimates the attacker's gains. 
It indeed considers a system that has no other protections or security system than the proof of work. 
Further, it does not consider that after a successful attack, the Bitcoin value is likely to plunge making it therefore unlikely for the attacker to spend her gain at current market value.
Finally, we should take into account that the attacker must have control over more than 50\% of the hashing power. 
This requires either huge investments in mining equipment (not taken into account in the formula) or other methods to control the mining farms, such as through a cyber or a conventional physical attack, which will also cost considerable amount of money. 
Therefore, we expect the parameter $p$ to be of the order of 1\% or less. 

Independently on the estimate of a realistic value for the parameter $p$, the principle that the cost of the proof of work must be a sizable fraction of the value transferred by the network to avoid double spending attacks should rest valid.
Specifically, according to this principle, we expect that, for a given system, the ratio between the cost of the proof of work and the value transferred by the network should oscillate around some constant value which reflects the fair balance between the possible gains in an attack and the cost to perform it.
In this paper, we test if this is indeed the case for the Bitcoin proof of work. 
For this purpose we are looking across the entire period of existence of Bitcoin, estimating the mining costs and comparing them with the value transferred through the network.
This is an amazing period during which the value transferred through the Bitcoin network has increased several million times and the hashing activity has increased by 10 orders of magnitude. Let us note that ten orders of magnitude is an immense change.
To put it into perspective this is the ratio between the diameter of the sun and the diameter of a one-cent coin. 
These are formidable changes to a scale never observed in financial systems or in human activity in general. 
We show in this paper that, despite these underlying formidable changes in the Bitcoin mining and trading activities, the ratio between the estimated mining cost and the transaction volume rests oscillating within a relatively narrow band supporting therefore the argument about the fair cost of the proof of work by \cite{aste2016fair}.

\begin{table}[htbp]
\centering
\begin{tabular}{|| c c ||} 
 \hline
 Start date & Bitcoin reward \\
 \hline
 3 Jan 2009 & 50.00 \\
 28 Nov 2012 & 25.00 \\
 9 Jul 2016 & 12.50 \\
 11 May 2020 & 6.25 \\
 \hline
\end{tabular}
\caption{Bitcoin reward per block mined.}
\label{table:reward}
\end{table}

\begin{table}[htbp]
\centering
\begin{tabular}{|| c c c c ||} 
 \hline
 Type & Hardware name & Date & J/Th \\
 \hline
 CPU & ARM Cortex A9 & 3 Oct 2007 & 877193 \\
 GPU & ATI 5870M & 23 Sep 2009 & 264550 \\
 FPGA & X6500 FPGA Miner & 29 Aug 2011 & 43000 \\
 ASIC & Canaan AvalonMiner Batch 1	& 1 Jan 2013 & 9351 \\
 ASIC & KnCMiner Jupiter & 5 Oct 2013 &	1484 \\
 ASIC & Antminer U1	& 1 Dec 2013 & 1250 \\
 ASIC & Bitfury BF864C55 & 3 Mar 2014 & 500 \\
 ASIC & RockerBox & 22 Jul 2014 & 316 \\
 ASIC & ASICMiner BE300 & 16 Sep 2014 & 187 \\
 ASIC & BM1385 & 19 Aug 2015 & 181 \\
 ASIC & PickAxe	& 23 Sep 2015 & 140 \\
 ASIC & Antminer S9-11.5 & 1 Jun 2016 & 98 \\
 ASIC & Antminer R4 & 1 Feb 2017 & 97 \\
 ASIC & Ebang Ebit 10	& 15 Feb 2018 & 92 \\
 ASIC & 8 Nano Compact & 1 May 2018 & 51 \\
 ASIC & Antminer S17 & 9 Apr 2019 & 36 \\
 ASIC & Antminer S19 Pro & 23 Mar 2020 & 30 \\
 \hline
\end{tabular}
\caption{Mining hardware with optimal energy efficiency and their dates of release.}
\label{table:hardware}
\end{table}

\section{Methodology}

\subsection{Estimation of the lower bound for the cost of Bitcoin mining}
The cost of Bitcoin mining is composed of three key elements: 1. the energy cost of mining; 2. the overheads for the maintenance of the mining farm such as infrastructure costs and cooling facilities; 3. the cost of purchasing and renewing the mining hardware. 
For the purpose of this study, we focus only on the first element, the energy cost of running the Bitcoin mining hardware which is likely to be the largest part of the cost and is the only one that can be estimated with some precision. 
The maintenance costs for running a Bitcoin mining farm varies widely depending on the location, design and scale of the facility and since such information are usually not disclosed to the public, it is infeasible to estimate it accurately. 
The sales price of mining hardware is publicly available but incorporating it into cost calculations is arduous because of the rapid rate of evolution in the industry and the information opacity regarding the market share of each hardware and the rate at which obsolete mining hardware are replaced. Newer mining hardware may achieve faster hash rates and higher energy efficiency but the renewing costs makes it unlikely that all Bitcoin miners immediately replace all their existing mining hardware with the the latest versions as they are released. Certainly a combination of both old and new mining hardware should coexist in the Bitcoin network as long as each machine continue to generate a profit. However, the market share of each hardware and its evolution over time is an unknown. 
With respect to the purpose of the present estimate of the lower bound of the mining cost, we must stress that the maintenance and the hardware costs must be anyway proportional to the energy consumption costs. 
By ignoring them we are under-estimating the total mining cost by some factor (presumably between 10\% to 20\%) but, beside this factor, the estimation of the overall behavior of the mining cost should not be significantly affected.

\subsection{Data}
Historic Bitcoin prices, the average hash rate, and the number of blocks per day from 2010 to 2020 were collected from \footnote{https://charts.bitcoin.com/btc} and the estimated total value of all transactions on the Bitcoin network was sourced from \footnote{https://www.blockchain.com/charts/}. 
Hash efficiency and other information regarding Bitcoin mining hardware were aggregated from manufacturer data and previous studies \citep{kufeouglu2019energy} and independently verified. 
Historical Brent Crude oil spot prices were collected from the United States EIA \footnote{https://www.eia.gov/dnav/pet/hist/rbrteD.htm}. 
The conversion rate between joules and barrels of oil equivalent, which may vary depending on oil grade, was set at  1 barrel = 5.54543 gigajoules based on the figures released by the BEIS. \footnote{https://assets.publishing.service.gov.uk/government/uploads/system/uploads/}

\subsection{Estimation of the energy costs of Bitcoin mining}
A mining hardware has an energy consumption that can be measured in joules per terahash (J/Th), and has a hashing speed that can be measured in terahashes per second (Th/s). 
For the purpose of estimating a lower bound to the energy costs of Bitcoin mining, we considered at any point in time that the entire network is adopting the most energy efficient machine available at that time. In situations where a mining hardware has different power setting options in which the user may choose to increase or decrease the hashing speed of the machine along with energy consumption, the most efficient power setting is used for calculation.

Most prior works has priced energy usage according to global average electricity prices (see for instance \cite{kufeouglu2019energy,derks2018chaining,vranken2017sustainability}). 
We have taken instead a slightly different approach by converting the energy consumption into barrels of oil equivalent and priced the oil according to the Brent Crude spot price. Indeed, the Brent Crude oil price is a publicly available daily value standardized around the world whereas electricity prices varies widely across different countries and suppliers and the actual price paid by the miner is not accessible.  
Note that there is a premium that electricity producers and distributors charge on the electricity price with respect to the oil cost and there can be also taxes. These extra charges depends on countries and situations but it is reasonable to figure out that they must add at least a 30\% extra to our estimate of the lower bound. However, coherently with our approach, we do not increase our figure by any assumed fraction, but we must bear in mind that our lower bound is certainly under-estimating the actual energy cost.

The lower bound of the energy costs of Bitcoin mining is estimated from total number of hashes times the energy cost of hashing by the most energy efficient Bitcoin mining hardware available on the market at any give time, divided by the conversion factor between energy and barrel of oil  and multiplyed by the cost of the oil.
Specifically, the lower bound for daily mining cost, $C_{t}$, is: 
\begin{equation}\label{TotalCostEstimate}
C_{t} = 
\frac{e_t H_{t}}{b}P_{t}
\end{equation}
where:\\
 $e_t$ is the energy efficiency of the most efficient mining hardware available on day $t$ in ($J/Th$) \\
 $h_{t}$ is the daily number of hashing operations in ($Th$) on day $t$;  \\
 $b$ is the joule of energy equivalent  per barrels of oil   $\approx 5.55 \times 10^9$ (J/barrel);  \\
 $P_{t} $ is the Brent crude oil spot price (USD/barrel) on day $t$. \\

\begin{figure}[ht]
\begin{center}
\includegraphics[width=0.7\linewidth]{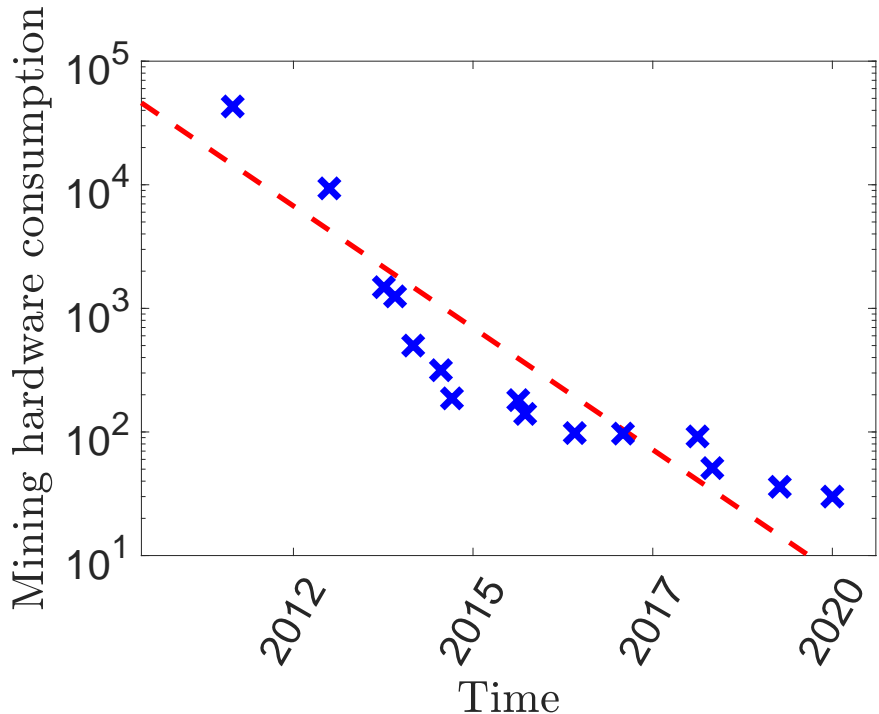}
\end{center}
\caption{Estimate of the lower bound for the energy consumption of the most efficient Bitcoin mining hardware, measured in J/Th. }
\label{2mining}
\end{figure}

\section{Results}

\subsection{Hardware efficiency variations}
Table \ref{table:hardware} reports a list of the Bitcoin mining hardware which consumed the least energy per  hash operations at the time of their release to the market. 
The improvement in energy efficiency for Bitcoin mining over time is reported in Figure \ref{2mining} where each symbol represents the most energy efficiency hardware at the respective time and the red dotted line displays a best-fit with an exponential curve: Energy Consumption $\sim c \exp(-\lambda t)$.
The best fit parameter is  $\lambda = 0.0025$, which means that hardware efficiency has been doubling every 10 months in average.
In a previous work a  power-law model was proposed by \cite{kristoufek}. 
However, the exponential model is more consistent with what is commonly expected for the rate of technology growth, according to the Moore's Law \citep{moore}.

\begin{figure}[ht]
\begin{center}
\includegraphics[width=0.7\linewidth]{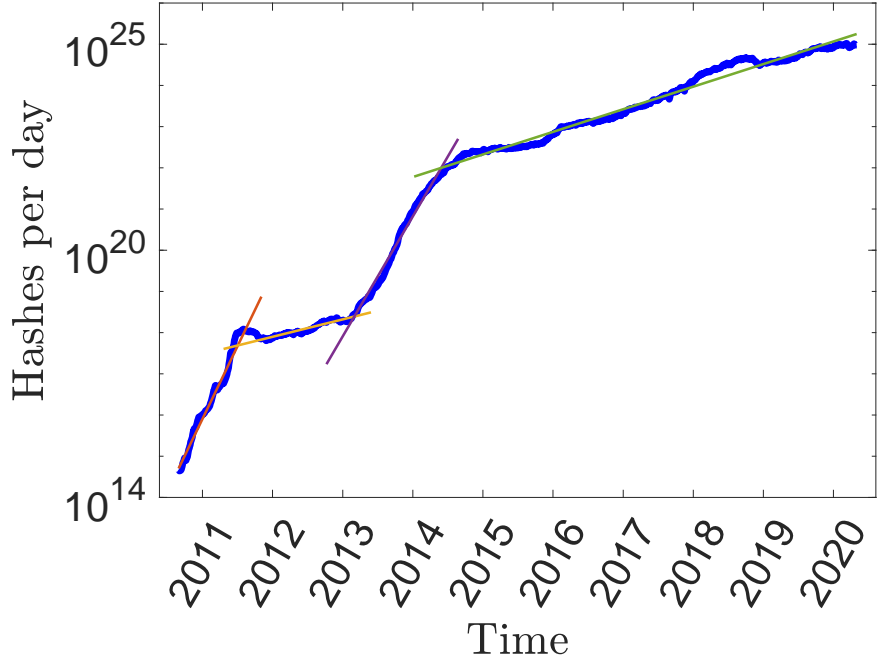}
\end{center}
\caption{Daily hashes computed by the Bitcoin network. The lines are best-fits with exponential growth laws in the corresponding sub-periods.
Doubling times are respectively (i) 33 days, during mid 2010 to mid 2011; (ii) 261 days, during mid 2011 to early 2013; (iii) 38 days during early 2013 to early 2015; (iv) 198 days, during early 2015 to early 2020.}
\label{hashes}
\end{figure}

\subsection{Hash computations variations}
Figure \ref{hashes} displays the total number of hashing operations per day. 
We note that the number of daily hashes have increased from $~10^{15}$  to $~10^{25}$ in the period between September 2010 to May 2020 when this paper was written. 
Daily hashes have been growing at exponential rates (linear trends in semi-log scale), which is in agreement with previous observations  \citep{dwyer}.
However, we can see from the figure that there are four, very distinct, periods with different grow rates.
Specifically:  (i) mid 2010 to mid 2011; (ii) mid 2011 to early 2013;  (iii)  early 2013 to early 2015;  (iv) early 2015 to early 2020.
The estimated best-fit doubling times in these periods are respectively: (1) 33 days; (ii) 261 days; (iii) 38 days; (iv) 198 days.

\begin{figure}[ht]
\begin{center}
\includegraphics[width=0.7\linewidth]{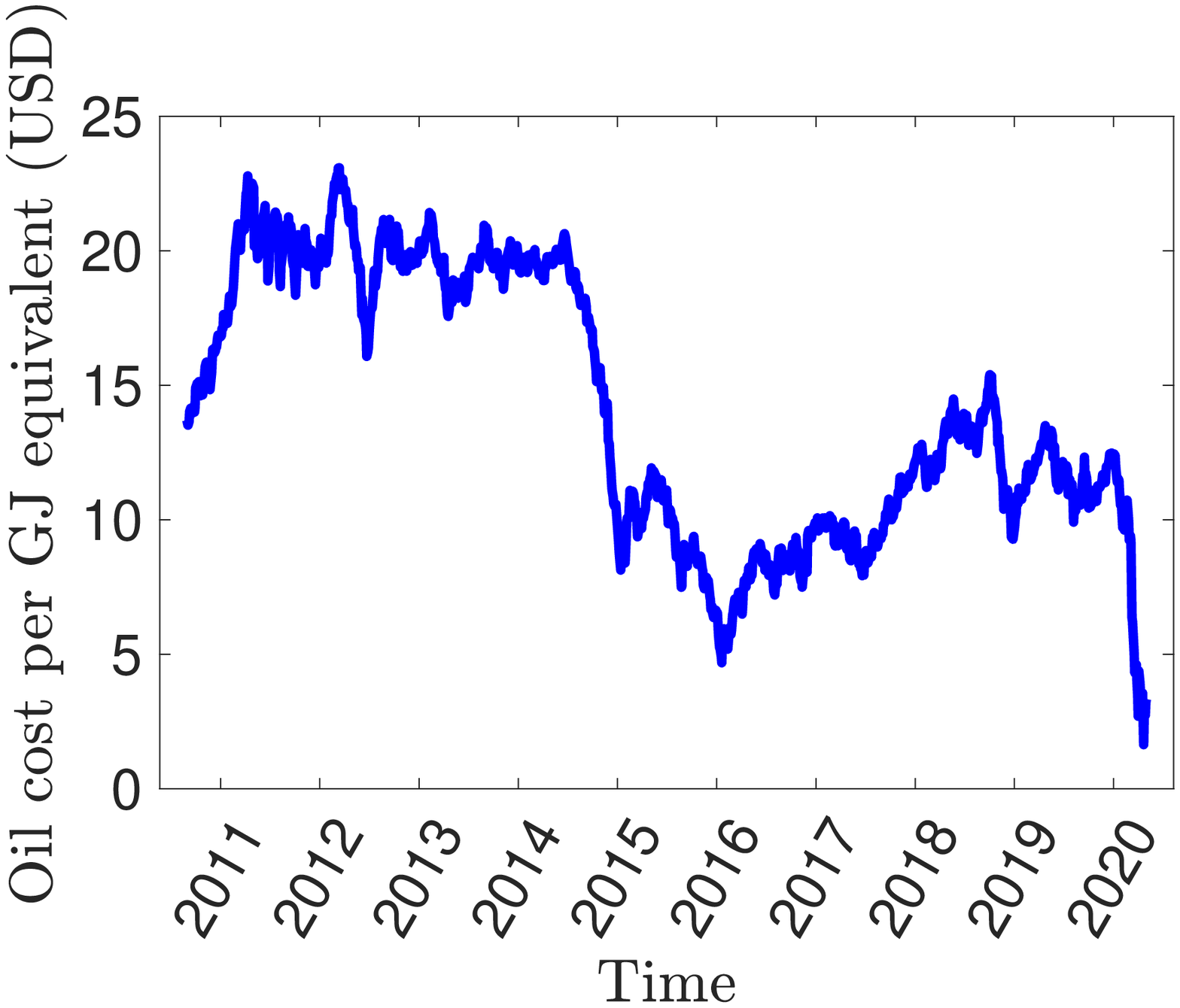}
\end{center}
\caption{Energy cost per gigajoule, measured in USD and converted from Brent Crude spot prices.}\label{energy}
\end{figure}

\begin{figure}[h!]
\begin{center}
\includegraphics[width=0.7\linewidth]{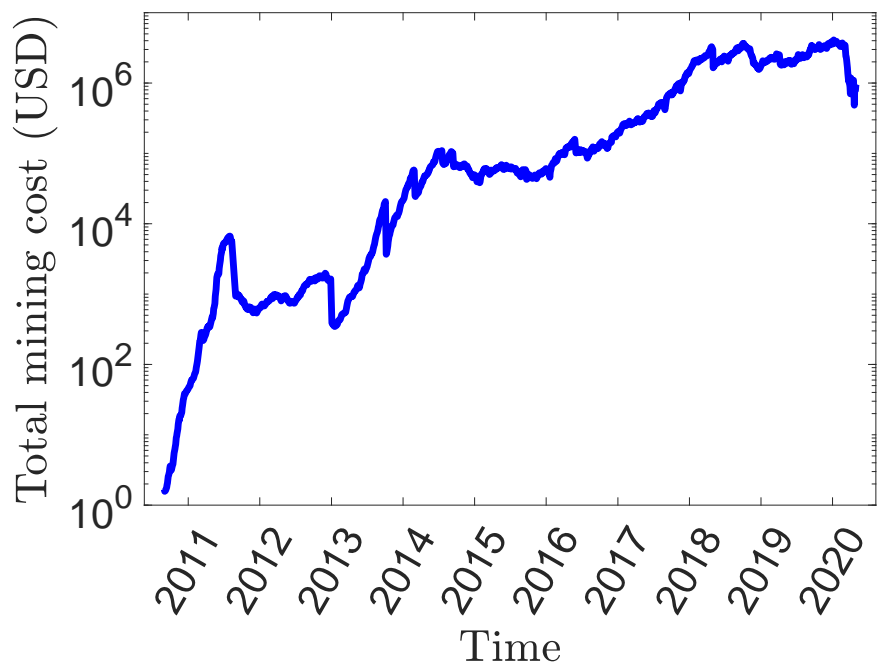}
\end{center}
\caption{Total daily mining cost $C_t$, reported in USD, estimated by using  Eq.\ref{TotalCostEstimate} }\label{valuecost}
\end{figure}

\subsection{Energy price variations}
Figure \ref{energy} shows the variations of the energy price per gigajoule in the period 2010 to 2020 computed from the Brent Crude spot prices. One can notice that the cost of one gigajoule of energy has two distinct levels - around 20 USD from 2011 to mid 2014 and around 10 USD from late 2014 to early 2020. Oil prices has since collapsed under the coronavirus pandemic, dropping to below 3 USD per gigajoule of energy. However, while large, the rate of change in energy price is several orders of magnitude smaller than the rate of change in the number of hashes. 

\begin{figure}[ht]
\begin{center}
\includegraphics[width=0.7\linewidth]{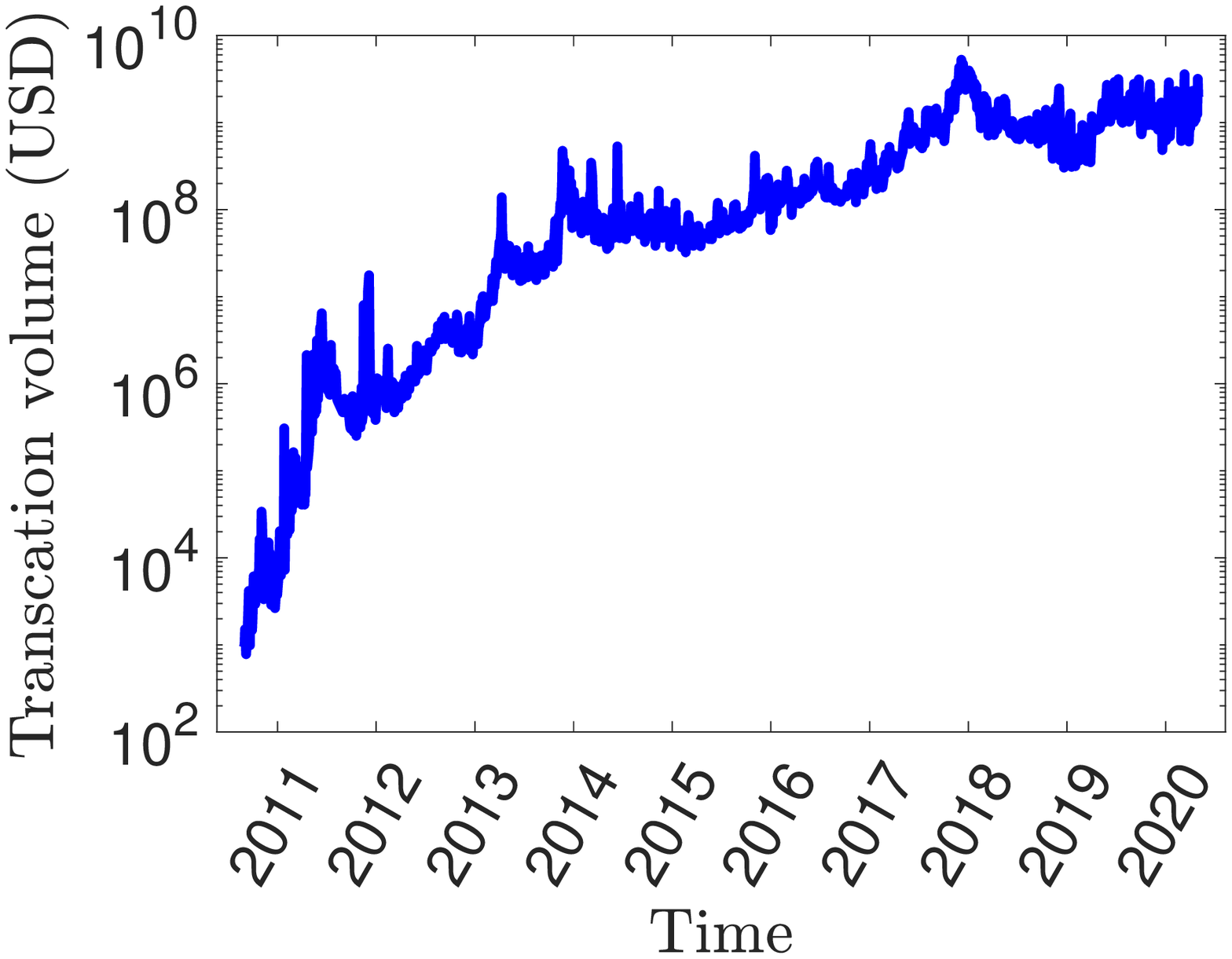}
\end{center}
\caption{Daily transaction volume $V_t$ reported in USD.}\label{value}
\end{figure}

\begin{figure}[h!]
\begin{center}
\includegraphics[width=0.7\linewidth]{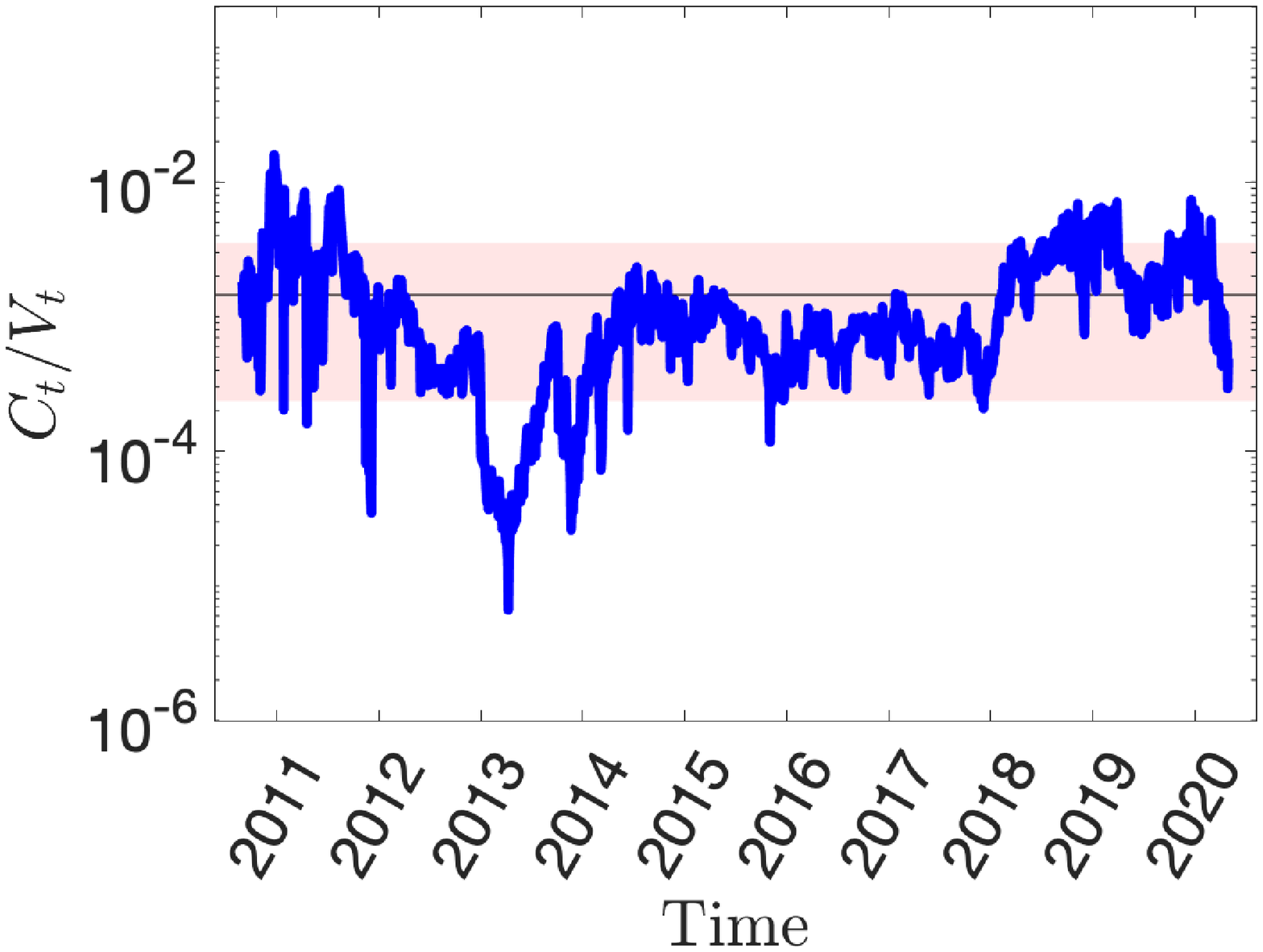}
\end{center}
\caption{
Ratio between the cost of mining and the total transaction volume $\frac{C_t}{V_t}$ on daily basis. 
The band is the region between the upper and lower 10\% quantiles and the line is the mean value, which is $0.14$\%. The mean value since 2018 (last plateau) is instead $0.3\%$.
}\label{ratio}
\end{figure}

\subsection{Lower bound mining cost estimate}
The lower bound of the total energy costs of Bitcoin mining is estimated as the minimum energy cost of each hash multiplied by the total number of hashes computed over a given period of time (a day in our case). 
Figure \ref{valuecost} reports the total mining daily cost in USD estimated by using Eq.\ref{TotalCostEstimate}, it varies from around 3 USD a day in 2010 to over 4 million USD a day in early 2020. 
Note that this is the lower bound estimate and the actual cost is presumably much larger.
The growth in mining costs is affected by both the changes in energy cost (see Fig.\ref{energy}) and by the increase in the hashing rate in the Bitcoin network (see Fig.\ref{hashes}). 
We note that the variations in energy cost oscillates in a much narrow band with respect to the changes in the daily number of hashes and therefore, the minimum Bitcoin mining costs (Figure \ref{valuecost}) mostly mirrors the growth in the total number of hashes.

\subsection{Transaction volume variations}
During the last ten years the Bitcoin network activity has also increased with increasingly larger amount of money transferred daily through the network. 
Figure \ref{value} reports the total transferred value per day in the Bitcoin network specified in USD.
One can see that the total daily volume of transactions has grown from about one thousand USD in 2010 to nearly one billion USD in 2020 for an increase by nine orders of magnitude.

\subsection{Ratio between mining cost and transaction volume}
Figure \ref{ratio} reports the ratio between the daily mining cost $C_t$ and daily transaction volume $V_t$. 
We observe that the ratio $\frac{C_t}{V_t}$ does not have any increasing or decreasing trend but rather is oscillating over the entire period from 2010 to 2020.
The largest variations  occurred in the first few years then, after 2014, the ratio value has stabilized into a plateau with then a jump to a higher plateau at the end of 2017 presumably due to the large decrease in Bitcoin price from over $19,000$ USD in December 2017 to just a little over $3,000$ USD in December 2018.
Despite the change in this relation between mining costs and transaction volume in 2017-18 and the change in Bitcoin prices in the same period, we note that in general this ratio is not correlated with the price of Bitcoin. There is actually a small negative correlation between the two for the daily variations.
Over the entire period, the mean value of $\frac{C_t}{V_t}$ is 0.14\% with bottom and top 10\% quantiles with values equal to 0.02\% and 0.4 \% respectively.
Note that this band of oscillation is within one order of magnitude whereas the underlying quantities $C_t$ and $V_t$ vary of nine orders of magnitude during the same period. 
If we limit our analysis to the last period after end 2017, we obtain a mean at 0.3\% and quantiles with values equal to 0.1\% and 0.4 \%.

\section{Discussion and Conclusions}

The proof of work allows a network of anonymous and untrustful parties to operate together without central authority control. 
It is a powerful instrument to keep a distributed system secure from malicious attacks. 
However, it has a high cost.
We estimate that presently at least a billion USD per year is burned by the Bitcoin network for the proof of work.
This amount corresponds to a one million times increase with respect to the costs in 2010.
However, although large, this amount is less than 0.5\% of the transaction volume over the network during the same period. 

Using data from 2009 to 2020, this paper quantifies the lower bound for the energy costs of Bitcoin mining and examines the relationship between this bound to the total value of transactions over time. 
We reveal that the ratio between mining cost and total transaction volume has not increased nor decreased over the last ten years despite Bitcoin mining activity having increased by ten billion times during the same period.   
Such an overall constant ratio is consistent with an argument, introduced by \cite{aste2016fair}, suggesting that such a ratio must be a sizable fraction of the  transaction volume and it corresponds to the minimum fraction that an attacker must double spend to make a profit (the quantity $p$ in Eq.\ref{eqn:fair}).  
Assuming the default block depth of $N=6$ in equation \ref{eqn:fair}, and using the last three years mean value for the fraction $\frac{C_t}{V_t}$  at $\sim 0.3\%$, we obtain that Bitcoin proof of work protects from a double spend attack at least a fraction around $p \sim 1.5\%$ of the global transaction volume. 
This result indicates that attacks that duplicate more than $\sim 1.5\%$  of the transaction volume could be profitable for the attacker because the cost of energy spent for the attack will be lower than the gain from the attack. 
This being a lower bound estimate that realistically could be an order of magnitude larger if all extra costs, beside the oil equivalent cost of mining energy, are included.  

We could therefore conclude that in the Bitcoin network the cost of proof of work is not at all too high. 
On the contrary it is actually too low to protect against double spending attacks. 
However, the proof of work is not the sole mechanism that provides protection of the Bitcoin network. 
The system also depends upon the high entry barriers  in terms of mining hardware and facilities costs. 
Further, Bitcoin value is built upon community trust so once a majority attack has been detected, the Bitcoin value is likely to collapse together with the potential attacker gains. 
This means that to launch a majority attack on Bitcoin, the attacker would thus not only need to invest substantial amounts of energy resources to gain over 50\% of the hashing power, they would also have to accept that many of the hardware costs incurred are unlikely to be recovered due to the inability of specialized ASICs to be repurposed for uses other than cryptocurrency mining and that even if the attack was successful, the value of Bitcoin would collapse so rapidly that there would be little economic gain. 
Finally, an attack involving a large fraction of the Bitcoin volume would be most likely detected by the network before its completion.  

Distributed systems and Blockchains can be secured through several other mechanisms that do not require computationally intensive proof of work. 
Indeed the proof of work is a mechanism introduced to produce qualified voters in a system of anonymous untrustful parties. 
Any mechanism that can verify identity of the voters’ or that can in any other way avoid uncontrolled duplications of the voters can reduce or eliminate completely the cost and even the need of a proof of work. 
However, these other mechanisms must relax also some other properties such as anonymity, openness or equalitarian distributed verification.

\section{Acknowledgments}
TA acknowledges support from ESRC (ES/K002309/1),  EPSRC (EP/P031730/1) and EC (H2020-ICT-2018-2 825215).

\section*{Author Contributions}
T.A. proposed the research, supervised and contributed to the data collection, performed the data analytics and co-drafted the paper. 
Y.S. collected, processed and analyzed the data, and co-drafted the paper. 
Both authors gave final approval for publication and agree to be held accountable for the content of the work.

\bibliographystyle{unsrt} 
\bibliography{references.bib}

\begin{thebibliography}{10}

\bibitem{aste2016fair}
Tomaso Aste.
\newblock The fair cost of bitcoin proof of work.
\newblock {\em Unpublished (2016); Available at SSRN 2801048}, 2016.

\bibitem{aste2017blockchain}
Tomaso Aste, Paolo Tasca, and Tiziana Di~Matteo.
\newblock Blockchain technologies: The foreseeable impact on society and
  industry.
\newblock {\em computer}, 50(9):18--28, 2017.

\bibitem{satoshi}
Satoshi Nakamoto.
\newblock Bitcoin: A peer-to-peer electronic cash system.
\newblock Technical report, Manubot, 2019.

\bibitem{bitcoinfirst}
Wing~Hong Chan, Minh Le, and Yan~Wendy Wu.
\newblock Holding bitcoin longer: The dynamic hedging abilities of bitcoin.
\newblock {\em The Quarterly Review of Economics and Finance}, 71:107--113,
  2019.

\bibitem{bitcoinmarketcap}
Klaus Grobys and Niranjan Sapkota.
\newblock Cryptocurrencies and momentum.
\newblock {\em Economics Letters}, 180:6--10, 2019.

\bibitem{marketcap}
Blockchain.
\newblock {Market Capitalization}.
\newblock 2020.
\newblock Accessed: 2020-01-21.

\bibitem{kufeouglu2019energy}
Sinan K{\"u}feo{\u{g}}lu and Mahmut {\"O}zkuran.
\newblock Energy consumption of bitcoin mining.
\newblock 2019.

\bibitem{derks2018chaining}
Jona Derks, Jaap Gordijn, and Arjen Siegmann.
\newblock From chaining blocks to breaking even: A study on the profitability
  of bitcoin mining from 2012 to 2016.
\newblock {\em Electronic Markets}, 28(3):321--338, 2018.

\bibitem{vranken2017sustainability}
Harald Vranken.
\newblock Sustainability of bitcoin and blockchains.
\newblock {\em Current opinion in environmental sustainability}, 28:1--9, 2017.

\bibitem{kristoufek}
Ladislav Kristoufek.
\newblock Bitcoin and its mining on the equilibrium path.
\newblock {\em Energy Economics}, 85:104588, 2020.

\bibitem{moore}
Gordon~E Moore.
\newblock Cramming more components onto integrated circuits.
\newblock {\em Proceedings of the IEEE}, 86(1):82--85, 1998.

\bibitem{dwyer}
Karl~J O'Dwyer and David Malone.
\newblock Bitcoin mining and its energy footprint.
\newblock 2014.

\end{thebibliography}

\end{document}